# System Virtualization and Efficient ID Transmission Method for RFID Tag Infrastructure Network

Shin-ichi Kuribayashi[1] and Yasunori Osana[1]

[1]Department of Computer and Information Science, Seikei University, Japan
E-mail: {kuribayashi, osana}@.st.seikei.ac.jp

*Abstract*

*The use of RFID tag which identifies a thing and an object will be expanded with progress of ubiquitous society, and it is necessary to study how to construct RFID network system as a social infrastructure like the Internet. First, this paper proposes the virtualization method of RFID tag network system to enable the same physical RFID network system to be used by multiple different service systems. The system virtualization not only reduces the system cost but also can dramatically reduce the installation space of physical readers and the operation cost. It is proposed that all equipments in the RFID network system except RFID tag could be shared with the conventional virtual technologies. Then, this paper proposes the conditional tag ID processing and the efficient tag ID transmission method which can greatly reduce the processing time and processing load in RFID tag Infrastructure network The conditional tag ID processing allows that tag ID is valid only at a certain time zone of day or in a certain area. The efficient tag ID transmission method uses the virtual network address of the service center as a part of the ID of an RF tag, which allows the direct ID forwarding to the service center.*

*Keywords*
*Virtual, RFID, tag, network*

## 1. Introduction

Radio Frequency Identification (RFID) is generic term for technologies employing radio waves for identifying objects. The advance in the information society is expanding the use of RFID tags [1]-[11]. RFID-based services are expected to become as essential to our lives as the Internet [2]. To bring this to reality, it is necessary to construct an RFID tag network system as a social infrastructure like electrical power, gas and water systems, which provides RFID tag service economically, safely and stably to enterprises, governmental organizations and people. Many efforts have been made to provide RFID tag services economically. For example, a common RFID tag network specification has been developed by standards bodies, such as EPC Global [12],[13] and Ubiquitous ID Center [13]. Universal readers that can support a variety of radio frequencies and interfaces have become available [13]. However, a universal RFID tag network system is yet to be developed.

The virtualization of RFID tag network system, which enables the same physical network system to be used by multiple different service systems, could be a promising technique to construct RFID network as a social infrastructure like the Internet. This virtualization not only reduces the system cost but also can dramatically reduce the installation space of physical readers and the operation cost. It is also expected that the RFID tag infrastructure network could offer the advanced features, which had not been offered in the conventional RFID tag network, in order to improve user convenience more than before. For example, it could useful to introduce the conditional tag ID processing in which tag ID is valid only at a certain time zone of day or in a certain area. This feature can eliminate the unnecessary processing at the server center and secure security.

Moreover, the types and volume of the identifiers (IDs) which need to be processed is likely to increase rapidly in an RFID network, compared to an existing network. Looking ahead to the widespread use of IDs for real time control in the future, it is clear that there will be a strong requirement for a reduction in the processing time related to the ID and in the processing load.

This paper assumes the EPC Global network architecture [12], and proposes the system virtualization method for RFID tag infrastructure network. This paper also proposes the conditional tag ID processing and the efficient tag ID transmission method for RFID tag infrastructure network. Section 2 discusses the basic concept of system virtualization of RFID tag network system for RFID tag infrastructure network, and proposes that all equipments in the RFID network system except RFID tag can be shared with the conventional virtual technologies for servers or networks. The association mechanism of a physical RFID tag and virtual reader is proposed. Section 3 proposes the conditional tag ID processing method which improves user convenience in RFID tag infrastructure network. Section 4 proposes the efficient tag ID transmission method that uses the virtual network address of the service center as a part of the ID of an RF tag, so allowing direct forwarding to the service center. Section 5 gives the conclusions. This paper is an extension of the study in [17],[18].

## 2. Basic Concept of Virtual RFID Tag Network System and Association Mechanism

### 2.1 Basic Concept of System Virtualization of RFID Tag Network System

A conventional RFID tag network has been built independently for each company or for each service system, as illustrated in Figure 1. This is based on the EPCglobal network architecture [9]. If an RFID tag network system is to become a social infrastructure, it is necessary to allow multiple service systems to share it. Sharing of an RFID tag network system can be achieved in two alternative ways:
  - Method I: Break usage time of RFID tag network system into timeslots and allocate different timeslots to different service systems.
  - Method II: Create, in each device, virtual elements, each for a specific service system.

This section proposes to adopt Method II because it is not realistic in Method I to allocate a fix timeslot to each application system.

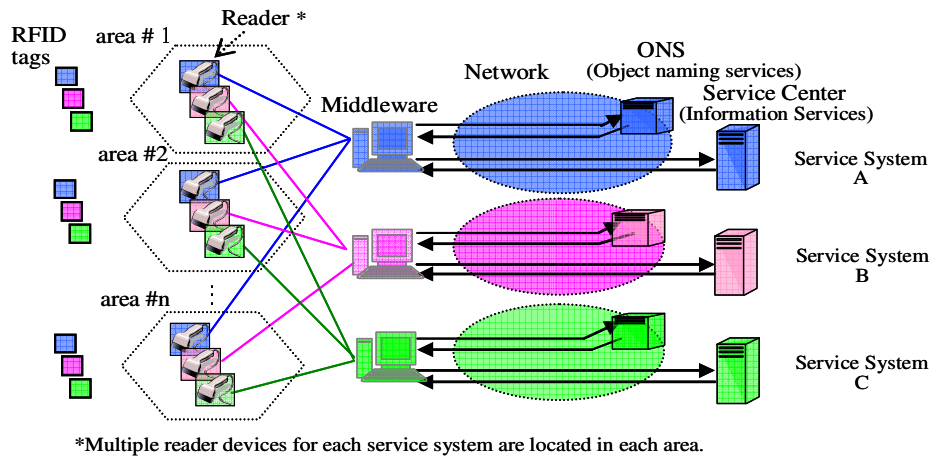

Figure 1. Existing RFID tag network system
(Each service system has its own physical network system)

*Multiple reader devices for each service system are located in each area.

Many technologies had been proposed for creating virtual networks or virtual systems [14],[15], and can be applied to the devices and network of an RFID tag network system. The universal reader which can read various types of RFID tags had been also proposed [13]. However, the method how to virtualize RFID tag, the construction method of virtual RFID tag network system by combining multiple virtual elements, the association mechanism of multiple virtual elements, and conditional tag ID processing function in shared RFID tag network system, had not been studied yet. The contribution of this paper is to provide the appropriate mechanisms to those issues.

First, it is proposed to use a different physical RFID tag for each service system. It is not appropriate to create virtual RFID tags because only one RFID tag is attached to each object, and the limit in functionality and memory space in an RFID tag makes sharing of the tag difficult. Figure 2 illustrates a construction image of the virtual RFID tag network system. Individual elements in the RFID tag network system, such as readers, middleware, ONS which translate tag IDs into addresses of service centers, network, and service center, are all virtualized. A virtual RFID tag network system is created by combining these virtual elements. It is assumed that the system management server centrally creates and manages the virtual elements. For example, if a virtual RFID tag network is to be built for service system B, the system management server first examines the requirements for service system B, and creates a virtual element in each of the service center, ONS, middleware and readers. It also creates related control data. Next, the system management server manages resource allocation to all elements in the RFID tag network system. It is noted that the system management server creates not just a single reader in the RFID tag network system but one reader in each of all areas covered by it. The virtual middleware contains the information about the association between it and each virtual reader. A virtual private network (VPN) is created in the network for each service system. The information about the association between a virtual element and the VPN is contained in the virtual element.

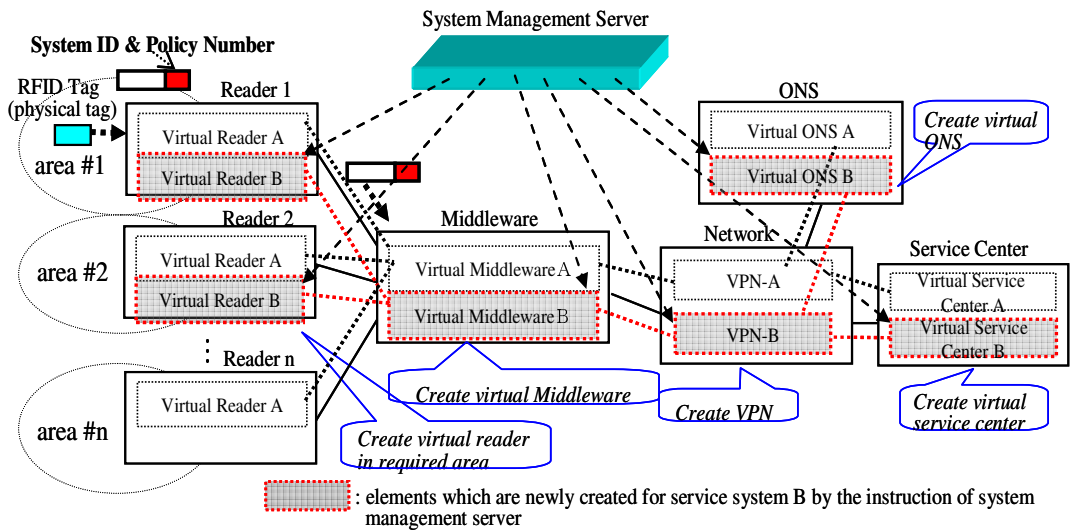

Figure 2. Construction image of a new virtual RFID tag network system
(The same physical network system is used by multiple different service systems)

Conversely, the system management server removes a virtual RFID tag network system for a specific service system by releasing the related virtual elements and VPN. It should be noted that virtualization of devices not only reduces the system cost through shared use of physical devices but also can dramatically reduce the installation space of physical readers and the operation of the RFID tag network system.

## 2.2 Association Mechanism between Physical RFID Tag and Virtual Reader

To build a virtual RFID tag network system, it is necessary to establish an association between physical RFID tags and virtual readers, as shown in Figure 3. Two alternative methods for this association can be conceived as illustrated in Figure 4:

<Method 1> Extract information required for identifying the service system from existing identifiers such as the tag ID stored in an RFID tag or the virtual reader identifier stored in a physical reader.

<Method 2> Introduce a new identifier (called "System ID") that identifies the service system and the processing policy. This identifier is used to maintain the association between RFID tags and a virtual reader, and to identify the required processing. The length of a system ID could be comparable to that of a domain manager, defined in EPCglobal sGTIN.

Method 2 is more desirable than Method 1 for the following reasons:
   1) There are such a wide variety of existing identifiers that it is difficult to take all of them into consideration in identifying the service system.
   2) Method 1 cannot support the case where a service is provided by multiple companies in collaboration. Although it may be possible to define a new corporate code for a group of companies working together, the number of potential combinations of companies is so large that this approach is not practicable.

International Journal of Computer Networks & Communications (IJCNC) Vol.2, No.6, November 2010

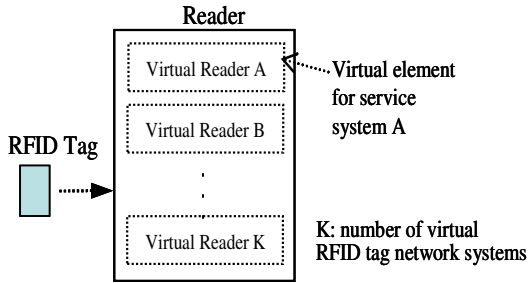

Figure 3. Relation between RFID tag and virtual reader element

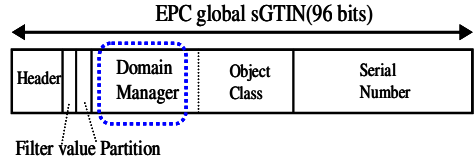

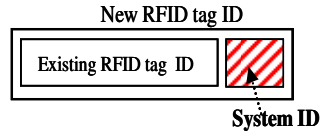

Figure 4. Methods to identify the relationship between RFID tag and virtual reader element

The system ID will be processed in the virtual RFID network system as follows (see Figure 2). First, the system ID, which is used to identify the service system that uses the particular RFID tag, is stored in the RFID tag together with its tag ID in advance. The reader that has read the system ID checks that there is a virtual reader associated with it, and sends the system ID to the middleware together with the tag ID when there is the associated virtual reader. The system ID is then handed over in succession to virtual network, ONS and finally to the virtual service center.  In this way, the system ID is recognized by the associated virtual elements regardless of which physical devices they are located in. Although the system ID is similar to the VPN-ID [14] (or VLAN-ID) in VPN, it is transferred separately from the VPN-ID.

## 3. Conditional Tag ID Processing Function

It is expected that the RFID tag infrastructure network could offer the advanced features, which had not been offered in the conventional RFID tag network, in order to improve user convenience more than before. We propose a conditional tag ID processing for this purpose. For example, a certain type of RFID tag is invalid at a certain time zone of day or in a certain area.   In the conventional RFID tag network system, tag ID is unnecessarily sent to the service center.   In the proposed method, tag ID can be discarded at the virtual reader and eliminate the unnecessary processing at the service center, as shown in Figure 5. There has been a mechanism for the middleware to accumulate a certain number of tag IDs before it sends them to the service center, our idea is that the virtual reader discards those tag IDs that are invalid at a certain time zone of day or in a certain area.

The following items should be taken into consideration to provide the proposed conditional tag ID processing:
 i) When an RFID tag is in a certain area, its tag ID is sent to the middleware.
 ii) When the virtual reader reads tag IDs at a certain time zone of day, it sends them to the

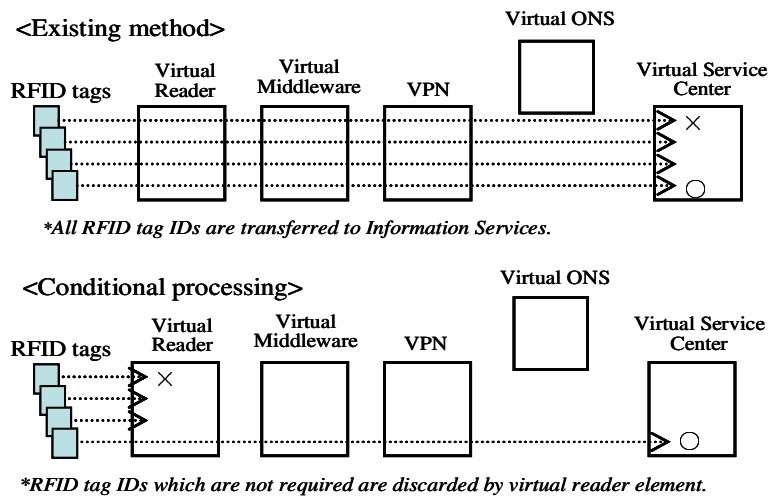

Figure. 5 Conditional tag ID processing method

middleware.
 iii) Even when the network is congested, certain specific packets are given priority and sent to the middleware.
 iv) The encryption system used by the RFID tag is identified.

The following three methods, illustrated in Figure 6, can be considered to implement the proposed method:
<Method A> Each virtual reader stores the information of conditions and actions for all tag IDs, and performs required actions based on the stored information.
<Method B> Multiple tag IDs are allocated to a single FRID tag.
<Method C> A new ID, 'Policy Number', is introduced that identifies the type of condition and processing required. Just as the system ID, the Policy Number is stored in the RFID tag.

Method C could be the most desirable for the following reasons:
 1) In Method A, the additional management capability should be implemented at all readers in the areas covered by the service system. It is supposed that the processing load for this implementation is heavy.
 2) In Method B, the memory space required in an RFID tag increases in proportion to the number of tag IDs.
 3) Method C does not affect existing RFID tags, and can be flexibly applied to different types of processing, even though it requires additional management of new number.

The introduction of Policy Number enables a different processing or encryption scheme to be selected depending on the area or the time zone of day, even for the same system ID. This method can be universally applied to all application systems, and simplify service development. After the Policy Number in RFID Tag has been read by a virtual reader, it is handed over in succession until it reaches the virtual service center. Each virtual device performs its processing based on the policy number. For example, only when the virtual reader has read a tag ID in an area other than the specified area, it sends the tag ID and system ID to the virtual middleware. Figure 7 provides a possible format and coding system of policy number.

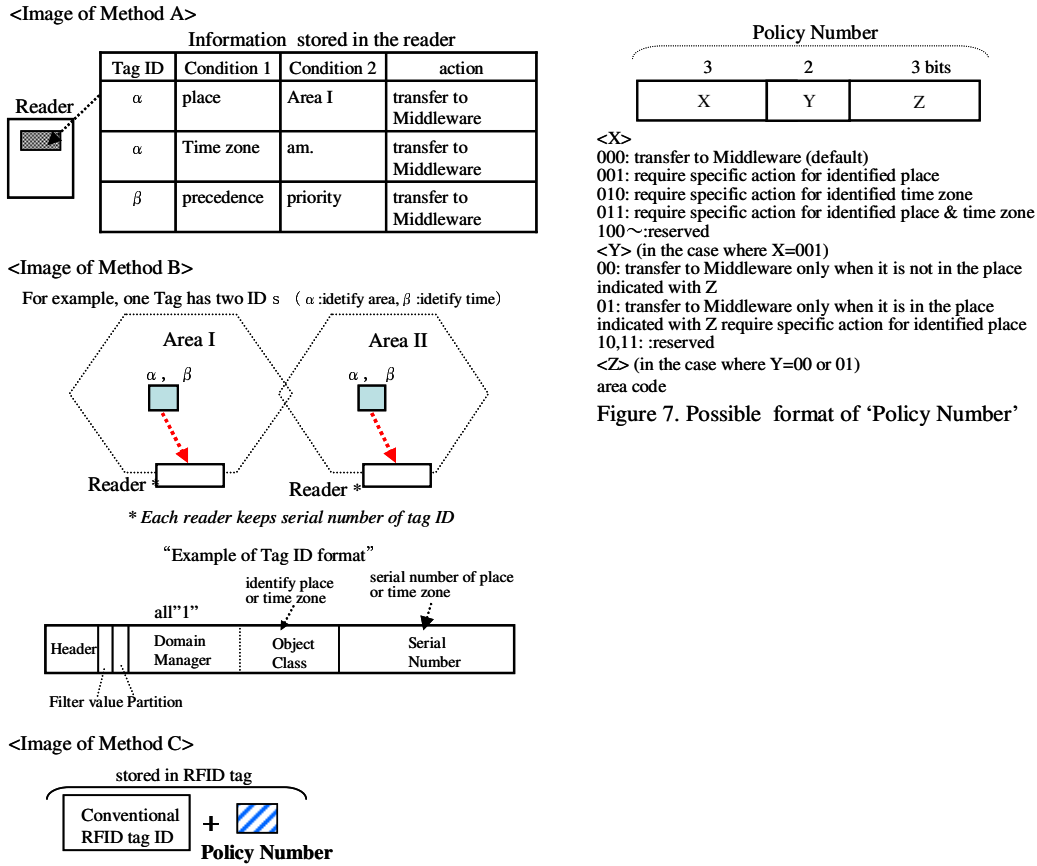

Figure 6. Methods for conditional tag ID processing

## 4. Proposed Efficient Tag ID Transmission Method

The types and volume of the identifiers (IDs) which need to be processed is likely to increase rapidly in an RFID network, compared to an existing network. Therefore, there is a strong requirement for a reduction in the processing time related to the ID and in the processing load.

The existing RFID network [12] requires the following two step process. In the first step, the terminal which reads the ID [13] of an RF tag accesses to a server in order to search for the appropriate network address of the service center. In the second step, it forwards the ID to the service center by using the acquired network address. The service center has the date related to the ID and executes a required processing with the data. Although the two step approach has the advantage that the ID value can be assigned independently of the network address, it results in an increase in both processing load and time to access the server. Use of a high performance server has been examined as a method of improving the situation. However, it is difficult to achieve a large reduction in access time and processing load because the need to access the

189

server to search for the network address remains. Moreover, it is difficult to adopt a method of caching because there are many types of ID that need to be handled in an RFID network. Therefore, we propose an efficient ID transmission method which can greatly reduce both the access time and processing load allows direct ID transmission to the service center.

The essential feature of the proposed method is that the network address of the service center is included as a part of the ID of an RF tag, in addition to System ID and Policy Number. This feature removes the need to access to another server in order to search for the network address of the service center which has the date related to the ID and executes a required processing with the data. In a word, it becomes possible to provide a direct ID transmission to the service center based solely on the value of the ID. VPI/VCI in ATM networks, MPLS tag in MPLS networks, MAC addresses in Ether networks, and IP addresses in IP networks, are candidates for the network address which is used as a part of the ID.

Figure 8 illustrates the image of the proposed ID transmission method in an RFID network. In Figure 8, an IPv4 address (private address) is used as a part of the ID. The network address of the service center, 192.168.1.0, is used as a part of the ID. The other part of the ID can be defined independently of the network address. When the reader reads the ID, the network address of the service center (192.168.1.0) is extracted and stored in the address part of an IP packet. The time and location of reading, as well as the entire ID, is set as the data part of the IP packet. The network node delivers the packet toward the service center using the network address stored in the address part of the IP packet. This is achieved by the normal routing function of the IP network. The service center that receives the IP packet extracts the necessary information in the data part and performs the necessary processing. For simplicity, Figure 8 does not consider the case where any information is sent back to the terminal from the service center.

The service center may be moved, or the function may be divided into several smaller servers. In that case, the network address of the service center needs to be changed and this requires changing the value of the ID which includes the network address. However, it is not practical to change all the related IDs in RFID tags. This paper proposes the following two techniques to avoid the above issue.

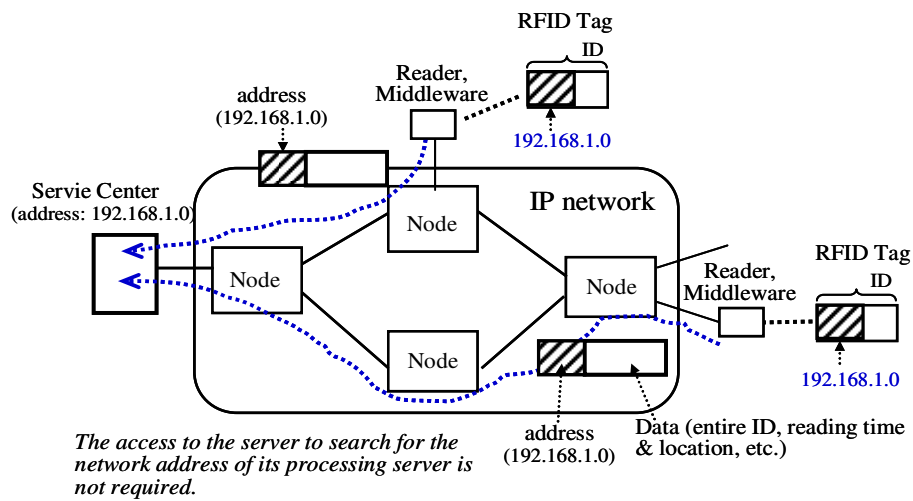

Figure 8. Processing image of the proposed ID transmission method in an RFID network

(1) Introduction of a 'virtual network address'

It is proposed that a virtual address is used as the network address of the service center (we call this a 'virtual network address' ). A virtual network address is assigned independently of the service center location and service center configuration. One virtual network address could have correspondence relationships with any physical service centers. A network node manages the relationship between virtual network addresses and the physical service center which accommodates these addresses. In this paper, the maintenance system is used to manage the relationship between a virtual network address and a physical service center in all network nodes. The virtual network address in RFID tag does not need to be modified when the relationship between a virtual network address and a physical service center is changed, even if one service center is divided into several smaller service centers.

Figure 9 illustrates one possible example of assigning virtual network addresses for apparel service centers. In this example, 6 virtual network addresses are assigned.

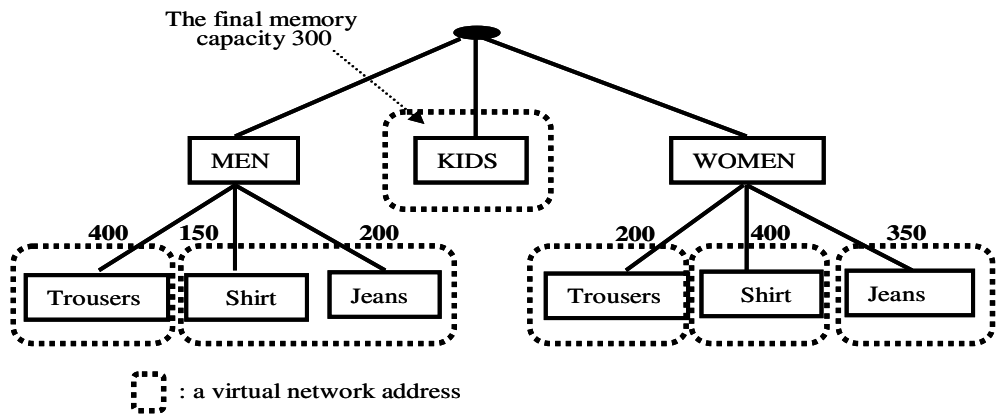

Figure 9. Example of assignment of virtual network addresses for Apparel service centers

(2) Introduction of 'packet re-transfer' function

It is difficult to change the data in all network nodes whenever the service center is moved or divided into several smaller servers. We propose to introduce a packet re-transfer function based on mobile IP [16]. Figure 10 illustrates the image of proposed method. The source node (node 1) transfers a packet toward the original destination node (node 2), which accommodated the service center before it was moved. Node 2 re-transfers the received packet to node 3 which now accommodates the service center after it has been moved. As a result, the ID which includes the virtual network address of the service center does not need to be changed.

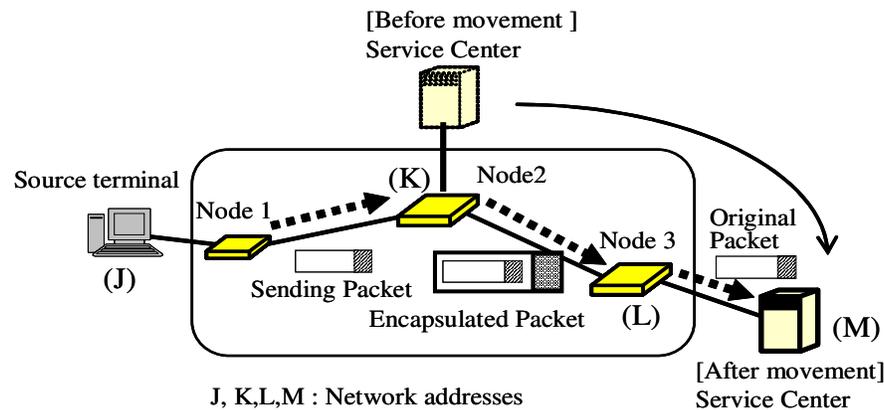

Figure 10. Packet transfer and re-transfer image in an RFID network

## 5. Conclusions

This paper has proposed the virtualization method of RFID tag network system to enable the same physical network system to be used by multiple different service systems. Specially, it has been proposed that all equipments in the RFID network system except RFID tag could be shared with the conventional virtual technologies for servers or networks. The proposed virtualization technique not only reduces the cost but also can dramatically reduce the space of reader installation and the operation cost.  It has been also proposed to add two new identifiers, in order to improve user convenience in the infra-structured RFID network system. One is 'System ID' to keep the association between a virtual reader element and a physical RFID tag, and the other is 'Policy Number' that specifies the conditional processing.

Then, this paper has proposed the conditional tag ID processing and the efficient tag ID transmission method which can greatly reduce the processing time and processing load in RFID tag Infrastructure network. The conditional tag Id processing allows that tag ID is valid only at a certain time zone of day or in a certain area The efficient tag ID transmission method uses the virtual network address of the service center as a part of the ID of an RF tag., which can allow the direct ID forwarding to the service center.

## Authors

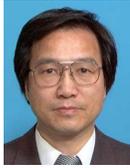

**Shin-ichi Kuribayashi** received the B.E., M.E., and D.E. degrees from Tohoku University, Japan, in 1978, 1980, and 1988 respectively. He joined NTT Electrical Communications Labs in 1980. He has been engaged in the design and development of DDX and ISDN packet switching, ATM, PHS, and IMT 2000 and IP-VPN systems. He researched distributed communication systems at Stanford University from December 1988 through December 1989. He participated in international standardization on ATM signaling and IMT2000 signaling protocols at ITU-T SG11 from 1990 through 2000. Since April 2004, he has been a Professor in the Department of Computer and Information Science, Faculty of Science and Technology, Seikei University. He is a member of IEEE, IEICE and IPSJ.

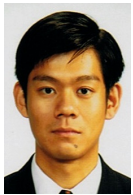

**Yasunori Osana** received the Ph.D. from the Graduate School of Science and Technology, Keio University in 2006. He is currently an assistant professor of the Department of Information and Computer Science, Seikei University. His research interests include architecture of network, computer and reconfigurable systems.